\documentclass[aps,prl,,twocolumn,groupedaddress]{revtex4}

\usepackage{graphics}
\usepackage{graphicx}

\usepackage{amsmath}

\def\fun#1#2{\lower3.6pt\vbox{\baselineskip0pt\lineskip.9pt
\ialign{$\mathsurround=0pt#1\hfil##\hfil$\crcr#2\crcr\sim\crcr}}}

\begin{document}

%{\Large \it Draft}
\preprint{Preprint JLAB-PHY-05-301}
\title{Two-Photon-Exchange Correction to Parity-Violating\\
Elastic Electron-Proton Scattering}

\date{February 14, 2005}
%%%                      author/address
\author{Andrei V. Afanasev$^{a)}$ and Carl E. Carlson$^{b)}$}
\affiliation{
$^{(a)}$Thomas Jefferson National Accelerator Facility, Newport News, VA 23606, USA\\
$^{(b)}$Department of Physics, College of William and Mary, Williamsburg, VA 23187, USA }

%                     Abstract
\begin{abstract}

Higher-order QED effects play an important role in precision
measurements of nucleon elastic form factors in electron scattering.
Here we introduce a two-photon exchange QED correction to the
parity-violating polarization asymmetry of elastic electron-proton
scattering.  We calculate this correction in the parton model using the formalism of
generalized parton distributions, and demonstrate that it can reach
several per cent in certain kinematics, becoming comparable in size with
existing experimental measurements of strange-quark effects in the
proton neutral weak current. 

\end{abstract}

\maketitle

%%%%%%%%%%%%%%%%%%%%%%%%%%%%%%%%%%%%%%%%%%%%%%%%%%%%%%%%%%%%%%%%%%%%%%%%%

%\section{Introduction}

%%%%%%%%%%%%%%%%%%%%%%%%%%%%%%%%%%%%%%%%%%%%%%%%%%%%%%%%%%%%%%%%%%%%%%%%%

{\it Introduction.} Recently the two--photon exchange mechanism in elastic electron--proton
scattering has attracted a lot of attention. The reason is that this
mechanism arguably accounts for the difference between the
high--$Q^2$ values of the electric to magnetic proton form factor ratio
$G_{Ep}/G_{Mp}$ \cite{TPE,Blunden} as measured in unpolarized (Rosenbluth)
and polarized electron scattering. Calculations of Ref.~\cite{Chen}
using a formalism of generalized parton distributions
(GPD's)~\cite{GPD} confirm the possibility, and decisive experimental
tests are under development~\cite{Brooks}.   See also
Ref.~\cite{deJager} for review of the problem.

The Rosenbluth/polarization controversy is resolved (putatively) by including
non-soft photon exchange into the box and cross-box diagrams of
electron-proton scattering. Since the additional exchanged photon is
not soft, it alters the spin structure of the electron--proton elastic
scattering amplitude and contributes to polarization observables~\cite{ABCCV}. 
These effects exceed spin-dependent corrections due to
bremsstrahlung~\cite{AAM}, because bremsstrahlung
photons can be constrained to be soft by experimental cuts.
The magnitude of the additional correction at fixed $Q^2$ depends
on the electron scattering angle, reaching a few percent. This result
motivated us to take a closer look at other electron scattering
measurements that require high precision.

In this note, we discuss the implications of two-photon exchange for
parity-violating electron scattering. The possibility of a radiative
correction of this type has been mentioned in the literature (see, {\it
e.g.}, Ref.~\cite{Musolf99}), but neither a general nor a model-based
analysis has been done so far. In our study, we derive expressions for
parity-violating observables in terms of generalized form factors, and
then present numerical results using the formalism of GPD's \cite{Chen}.
We conclude that the two-photon exchange mechanism introduces a
systematic correction to parity-violating electron-proton scattering
asymmetry at a level of one or more per cent, which is comparable in size with
existing experimental constraints of the strange quark contribution to this
asymmetry.
 
%%%%%%%%%%%%%%%%%%%%%%%%%%%%%%%%%%%%%%%%%%%%%%%%%%%%%%%%%%%%%%%%%%%%%%%%%

%\section{Currents and Observables}

%%%%%%%%%%%%%%%%%%%%%%%%%%%%%%%%%%%%%%%%%%%%%%%%%%%%%%%%%%%%%%%%%%%%%%%%%

\begin{figure}
\includegraphics[width=7cm]{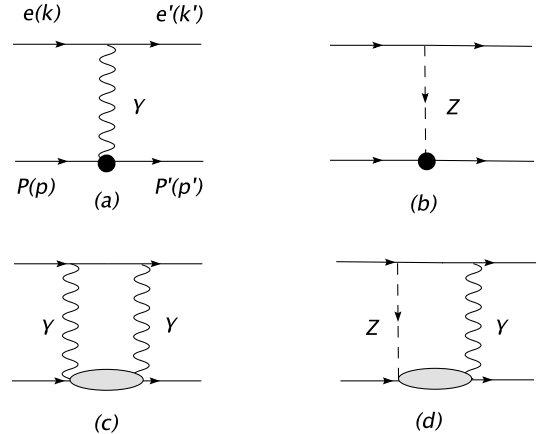}
\caption{Diagrams of Born approximation (a, b), two-photon exchange (c) and $\gamma Z$ box (d)
for elastic e-p scattering in a Standard Model of electroweak interactions.
Corresponding cross-box diagrams are implied.}
\label{diagrams}
\end{figure}

{\it Currents and Observables.} At tree level, parity-violation in
electron scattering is caused by interference of the diagrams with
single photon and single $Z^0$-boson exchange shown in
Fig.~\ref{diagrams}(a) and (b). Leading-order radiative corrections
include, but are not limited to, the box diagrams of 
Fig.~\ref{diagrams}(c,d), where the blobs between photon and Z-boson
coupling to the nucleon denote excitations of all possible intermediate
states by the electromagnetic and weak neutral currents. Therefore
calculation of such corrections depends on nucleon structure and
requires modeling of the nucleon radiative weak amplitude and Compton
amplitude under the loop integral. The mechanism of
Fig.~\ref{diagrams}(d), the $\gamma Z$-box, contributes to
parity-violating electron-nucleon interaction through interference with
one-photon exchange Fig.~\ref{diagrams}(a). It was calculated  by
Marciano and Sirlin~\cite{Marciano:1983ss} for the case of atomic
parity violation corresponding to the limit of zero momentum transfer.

%%%%%%%%%%%%%%%%%%%%%%%%%%%%%%%%%%%%%%%%%

%%%%%%%%%%%%%%%%%%%%%%%%%%%%%%%%%%%%%%%%%

Let us consider one more possibility, namely, the radiative correction
due to interference between the diagrams with $Z$-boson exchange
Fig.~\ref{diagrams}(b) and two-photon exchange Fig.~\ref{diagrams}(c).
This mechanism of $2\gamma\times Z$-interference contributes to the
observed asymmetry at the same order $\cal O (\alpha)$ as the $\gamma
Z$ box discussed in Ref.~\cite{Marciano:1983ss}; it comes in addition to
the parity-conserving correction to the unpolarized cross section from
two-photon exchange. For purely leptonic processes, the contribution of
the box diagrams is calculable exactly as a part of the leading-order
electroweak radiative correction~\cite{Bohm:1986rj}. However,
additional knowledge of hadronic structure is needed to calculate a
similar correction for semi-leptonic weak processes. A partial solution
of this problem was introduced by Mo and Tsai in the classic
paper~\cite{MoTsai} on radiative corrections to electron-proton
scattering, in which the photon in diagram  Fig.~\ref{diagrams}(c)
was treated in an infra-red approximation, thereby reducing the
hadronic-structure dependence to Born-approximation form factors. We
should note that if the infra-red approximation is used, then the
combined contribution of the boxes Fig.~\ref{diagrams}(c,d) would
factor out and exactly cancel in the polarization asymmetries. To
obtain the correction to polarization asymmetries, one needs to go
beyond the infra-red approximation for two-photon exchange, and this is
the goal of the present article.

Let us first present general formulas for the electromagnetic and weak
currents and the parity-violating asymmetry. The Born matrix element of
Fig.~\ref{diagrams}(a) is described in a standard way in terms of
proton electromagnetic form factors $F_{1,2p}^\gamma$. The exchange of the
$Z$-boson Fig.~\ref{diagrams}(b) is parameterized at tree level as
\cite{Kaplan88}			
\begin{eqnarray}		\label{BornCurrents}
{\cal M}^Z&=& -\frac{G_F}{\sqrt{2}} j_{\mu}^Z J_{\mu}^Z, \\
j_{\mu}^Z&=&\bar{u}(k') \gamma_{\mu} (g_V^e-g_A^e\gamma _5) u(k),\\
J_{\mu}^Z&=&\bar{u}(p') \Bigl[ \gamma_{\mu} F_1^Z-\sigma_{\mu\nu}q_{\nu} \frac{F_2^Z}{2 M}+
\gamma_{\mu}\gamma_5 G_A^Z\Bigr] u(p),
\end{eqnarray}
where $G_F$ is Fermi constant, $F^Z_{1,2p}$ are the form factors of the proton neutral weak current, 
$M$ is the proton mass, and $q=k-k'$.
In the Standard Model of electroweak interactions,
the weak form factors $F^Z_{1,2p}$ are related to the proton and neutron
electromagnetic form factors by the expression:
\begin{eqnarray}
F^Z_{1,2p}(Q^2)= (1-4\sin^2\theta_W) F_{1,2p}^{\gamma}-F_{1,2n}^{\gamma}-F_{1,2}^{s},
\end{eqnarray}
where $\theta_W$ is the weak mixing (Weinberg) angle and the quantities $F_{1,2}^{s}$ denote contributions to the
neutral weak current from strange (and heavier) quarks. Other combinations
of form factors are also used, namely: $G_E=F_1-\tau F_2$, $G_M=F_1+F_2$, where $\tau={Q^2}/(4M^2)$.

The axial form factor is
\begin{equation}\label{GAZ}
G_A^Z(Q^2)=-\tau_3 G_A(Q^2)+\Delta s,
\end{equation}
where $\tau_3$= +1(--1) for protons (neutrons) and $\Delta s$ stands
for the strange-quark contribution. The isovector axial form factor  is
normalized at $Q^2$= 0 to the neutron $\beta$-decay constant as
$G_A(0)=-g_A/g_V= + 1.2670\pm 0.0035$. Note that all the above
expressions are valid at tree level, when higher-order radiative
corrections are neglected.

Interference between photon and $Z$-boson exchange produces an
experimentally observable parity-violating asymmetry $A_{PV}$ in the
scattering of longitudinally polarized electrons on an unpolarized
target~\cite{ApvSM}. It can be expressed in terms of the above quantities as 
\begin{equation}\label{asymBorn}
A^{Born}_{PV}=\frac{\sigma_R-\sigma_L}{\sigma_R+\sigma_L}=
-\frac{G_F Q^2}{4\pi\alpha\sqrt{2}}\frac{A^{Born}_E+A^{Born}_M+A^{Born}_A}{\bigl[\epsilon (G^{\gamma}_{Ep})^2+
\tau (G^{\gamma}_{Mp})^2\bigr]},
\end{equation}
where
\begin{eqnarray}\label{ff2as}
A^{Born}_E&=&-2 g_A^e\epsilon G_{Ep}^Z G_{Ep}^{\gamma},\ A^{Born}_M=-2 g_A^e\tau G_{Mp}^Z G_{Mp}^{\gamma} 
\nonumber \\
A^{Born}_A&=&2 g_V^e\sqrt{\tau (1+\tau) (1-\epsilon^2)} G_A^Z G_{Mp}^{\gamma}  \\
\epsilon&=&(1+2 (1+\tau)\tan^2\frac{\theta_e}{2})^{-1},\nonumber
\end{eqnarray}
and where $g_V^e=-(1-4\sin^2\theta_W)/2,\ g_A^e=-1/2$ are the electron
coupling constants from Eq.~(\ref{BornCurrents}).

Let us now consider the case when electron--proton scattering is caused
by exchange of more than one photon. Neglecting the electron mass, the
electron--nucleon scattering amplitude can be presented in the form
described by three complex scalar invariants, see also \cite{TPE,Chen}:
\begin{eqnarray}
\cal M^{\gamma}&=&\frac{e^2}{Q^2}\Bigl[\bar{u}(k') \gamma_{\mu} u(k)\nonumber \\ 
&\times&\bar{u}(p')\bigl(\gamma_{\mu} F'_{1}-\sigma_{\mu\nu}q_{\nu} \frac{F'_2}{2 M}\bigr) u(p)\\
&+&\bar{u}(k') \gamma_{\mu}\gamma_5 u(k)\times \bar{u}(p')\gamma_{\mu}\gamma_5 G'_A u(p)\Bigr].\nonumber
\end{eqnarray}
The invariants, or generalized form factors, may be separated into parts coming from one-photon exchange, Fig.~\ref{diagrams}(a),
and parts from two- or more-photon exchange, Fig.~\ref{diagrams}(c),
\begin{eqnarray}\label{FFs}
G'_M(\epsilon,Q^2)&=&G^{\gamma}_M(Q^2)+\delta G'_M(\epsilon,Q^2) \nonumber \\
G'_E(\epsilon,Q^2)&=&G^{\gamma}_E(Q^2)+\delta G'_E(\epsilon,Q^2) \\
G'_A(\epsilon,Q^2)&=&\delta G'_A(\epsilon,Q^2),	\nonumber
\end{eqnarray}
where $G^{\gamma}_M(Q^2)$ and $G^{\gamma}_E(Q^2)$ are the usual
magnetic and electric form factors that parameterize matrix elements of
nucleon electromagnetic current, and the corrections of order $\cal
O(\alpha)$ from multi--photon exchange are represented by the
quantities $\delta G'_M$, $\delta G'_E$, and $G'_A$. 

Equipped with the corrected expression for the proton electromagnetic
amplitudes, it is straightforward to update the formula for the
parity-violating asymmetry $A_{PV}$ which properly includes the
two-photon box contributions. The asymmetry takes the form,
\begin{equation}\label{newApv}
\begin{split}
&A_{PV}= -\frac{G_F Q^2}{4\pi\alpha\sqrt{2}}\times \\
&\frac{A_E+A_M+A_A+A'_M+A'_A}{\epsilon |G'_{Ep}|^2+\tau |G'_{Mp}|^2
+2 \sqrt{\tau (1+\tau )(1-\epsilon^2)} G^{\gamma}_{Mp} \Re(G'_{Ap})},
\end{split}
\end{equation}
where the formulas for $A_{E,M,A}$ are obtained from the corresponding
formulas for $A^{Born}_{E,M,A}$ of Eq.~(\ref{ff2as}) by replacing
electromagnetic form factors $G^{\gamma}_{E,M}$ with real parts of
their generalized versions $\Re(G'_{E,M})$. 
Two additional terms, $A'_M$ and $A'_A$, arise from interference
between the axial-like terms in the two-photon exchange amplitude
and $Z$-boson exchange,  
\begin{eqnarray}\label{newFF2a}
A'_A&=&2 g^e_V (1+\tau) G_{A}^Z \,\Re(G'_{Ap}) \\
A'_M&=&-2 g^e_A \sqrt{\tau (1+\tau) (1-\epsilon^2)} G_M^Z 
		\,\Re(G'_{Ap}). \nonumber 
\end{eqnarray}

One can see from Eq.~(\ref{newFF2a}) that the parity-violating
asymmetry acquires new terms from the modified spin structure of
electron--proton scattering amplitude coming from two-photon exchange.
Because of these terms, the updated expressions for $A_{PV}$ cannot be
obtained by just updating the nucleon electromagnetic form factors,
but also require including this new $\cal O(\alpha)$ correction to the
observed asymmetry.

Thus the two-photon exchange correction to $A_{PV}$ is a combined
effect from the correction to the unpolarized cross-section in the
denominator of Eq.~(\ref{newApv}) and the correction to the
polarization-dependent numerator. To the leading order in
(electromagnetic) $\alpha$, the former is due to interference between
one- and two-photon exchange diagrams Fig.~\ref{diagrams}(a) and (c),
whereas the latter arises due to interference of $Z-$boson and
two-photon exchange Fig.~\ref{diagrams}(b) and (c). 
%Results of
%calculations for both these effects are presented in the following
%section. 

%%%%%%%%%%%%%%%%%%%%%%%%%%%%%%%%%%%%%%%%%%%%%%%%%%%%%%%%%%%%%%%%%%%%%%%%%

%\section{GPD Calculation of two-photon exchange correction to $A_{PV}$}

%%%%%%%%%%%%%%%%%%%%%%%%%%%%%%%%%%%%%%%%%%%%%%%%%%%%%%%%%%%%%%%%%%%%%%%%%

{\it GPD Calculation of two-photon exchange correction to $A_{PV}$.} We
evaluate the contribution of two-photon exchange to the
parity-violating asymmetry using a partonic formalism.  The calculation
of the generalized form factors given in Eq.~(\ref{FFs}) is described
in detail in Refs.~\cite{Chen,ABCCV}, and here we will only briefly
remind the reader of the main steps.

For the 2-photon exchange mechanism of Fig.~\ref{diagrams}(c), a
`handbag' approximation is used, in which the two photons are coupled
directly to a point-like quark. The photon phase space in the
4-dimensional loop integral is separated into `soft' and `hard' regions
using the gauge-invariant procedure of~\cite{grammer}. The `soft' part
is treated separately in accordance with a low-energy theorem, while
the `hard' part calculation is performed using GPD's to describe
the emission and reabsorption of a quark by a nucleon.  The same
parameterization is used for the GPD's as in Ref.~\cite{Chen}, with
constraints on $x$ (quark momentum fraction) and $Q^2$ dependence coming from available data on
elastic nucleon form factors and parton distribution functions of
inclusive deep-inelastic scattering. The calculation is not extended to
small values of $\epsilon$ because the model~\cite{Chen} in its present
form has limited applicability in this region, which corresponds to
small values of Mandelstam variable $-u<M^2$. The parton model approach
used here also restricts 4-momentum transfers to the region $Q^2>M^2$.

\begin{figure}[h]
\includegraphics[width=7cm]{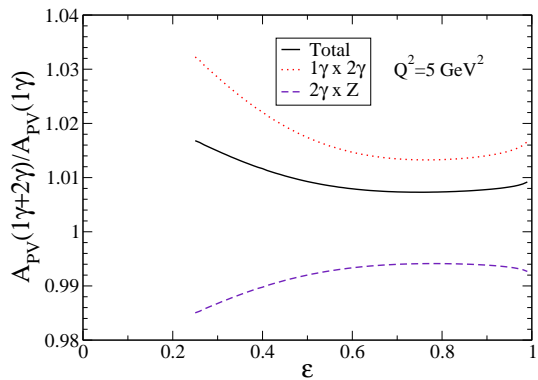}
\caption{Two-photon exchange correction to parity-violating asymmetry as a function of
$\epsilon$
at $Q^2$= 5 GeV$^2$. Also shown are separate effects from the parity-conserving
 $1\gamma\times 2\gamma$-interference 
and parity-violating $2\gamma\times Z$-interference.}
\label{CorPlot}
\end{figure}

\begin{figure}[h]
\includegraphics[width=7cm]{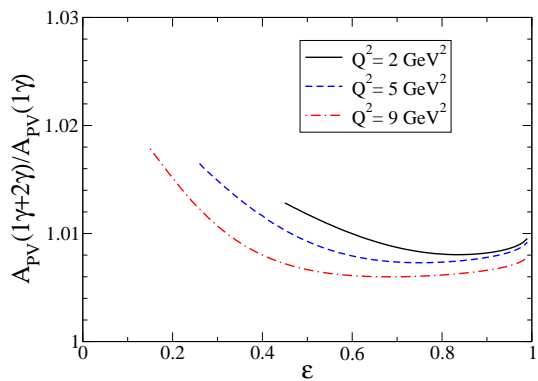}
\caption{$Q^2$-evolution of two-photon exchange correction to parity-violating asymmetry 
for several values of $Q^2$= 2 GeV$^2$ (solid curve), 5~GeV$^2$ (dashed curve) and 9 GeV$^2$
(dash-dotted curve).}
\label{CorVsQ2}
\end{figure}

The results of the calculation are shown in Figs.~\ref{CorPlot}
and~\ref{CorVsQ2} for the ratio 
\begin{equation}
R=\frac{A_{PV}(1\gamma+2\gamma)}{A_{PV}(1\gamma)},
\end{equation}
where $A_{PV}(1\gamma)$ is the Born asymmetry of Eq.~(\ref{asymBorn})
and $A_{PV}(1\gamma+2\gamma)$ is the calculation with two-photon
exchange correction included. It is instructive to compare
contributions to the above ratio from the polarization-dependent
numerator and the polarization-independent cross-section in the denominator
of Eq.~(\ref{newApv}), because they come from interference of different
pairs of diagrams. The two-photon exchange correction reduces the
magnitude of both the denominator and the numerator of
Eq.~(\ref{newApv}) by several per cent, leading to partial cancellation
of the effect for the asymmetry and increasing the magnitude of $A_{PV}$
compared to its Born value. 
\bigskip

%%%%%%%%%%%%%%%%%%%%%%%%%%%%%%%%%%%%%%%%%

%%%%%%%%%%%%%%%%%%%%%%%%%%%%%%%%%%%%%%%%%

%%%%%%%%%%%%%%%%%%%%%%%%%%%%%%%%%%%%%%%%%

%%%%%%%%%%%%%%%%%%%%%%%%%%%%%%%%%%%%%%%%%

The correction is both $Q^2$ and $\epsilon$ dependent. It is increasing
toward backward electron scattering angles (small $\epsilon$), reaching
above one per cent. Quantitatively, a 1 per cent increase in the magnitude of
$A_{PV}$ due to two-photon exchange results in about the same
percentage decrease in the magnitude of the extracted $G_{Mp}^Z$, 
so that the extracted strange $G^s_M$ would become more positive by about 1\% of the value
of $G_{Mp}^Z$. It would have an even larger impact were one extracting $\Delta s$ at
backward angles, changing the extracted $\Delta s$ by about 10\% the
value of $G_A(Q^2)$, Eq.~(\ref{GAZ}). The magnitude
of the correction in the model is therefore comparable in size with
experimentally measured contributions (albeit done at lower $Q^2 < 1$
GeV$^2$) to the scattering asymmetry from strange quarks in the proton
neutral weak current \cite{PVES_exp}; see also Ref.~\cite{Beise03} for 
review of the current experimental status.

Our result also calls for update of the model corrections to the proton
neutral weak current from the $\gamma Z$-box  and an extension of model
calculations to the region of lower momentum transfer $Q^2<$ 1 GeV$^2$.

Let us discuss the possibilities of setting experimental constraints on
two-photon exchange and $\gamma Z$ corrections to $A_{PV}$. It is known
that the difference between the cross sections for electron and
positron electromagnetic scattering on a nuclear target provides a
direct measure of the two-photon exchange contribution to the cross
section~\cite{Mar68}. However, the situation is more complicated for
weak currents, for which the vector and axial-vector terms change
relative sign under charge conjugation.  As a result, the
parity-violating asymmetry at tree level can be presented in the form:
\begin{equation}\label{elposApv}
A_{PV}^{\pm}\propto g^e_V \alpha_A\pm g^e_A \alpha_V,
\end{equation}
where the upper (lower) sign corresponds to scattering posi\-trons
(electrons) and the quantities $\alpha_V$ ($\alpha_A$) are due to the
vector (axial-vector) hadronic neutral weak current interfering with
the electromagnetic current. With two-photon exchange and $\gamma Z$
contributions included, each term in the above expression
(\ref{elposApv}) receives a correction $\delta\alpha_{A,V}$ that has
opposite signs for electrons and positrons, {\it i.e.}
$\alpha_{A,V}\to\alpha_{A,V}\pm\delta\alpha_{A,V}$. As a result, a
comparison of electron vs. positron scattering on a nucleon target 
is not sufficient for separating out the contributions with an
extra photon exchange between the lepton and the hadron. However, there
is an exception for specific quantum numbers of the target. For
spin-zero targets, ({\it e.g.}, $^4He$) and elastic scattering, the
hadronic weak current has only a vector component and thus only the term
$\propto g^e_A$ remains in Eq.~(\ref{elposApv}). Therefore for such a
target, the sum of asymmetries from positrons and electrons,
$A_{PV}^{+}+A_{PV}^{-}$, is proportional to the effects with an extra
photon exchange shown in Fig.~\ref{diagrams}(c,d). Such an experiment,
however, is difficult to implement at this time without
high-current polarized positron beams.

As far as the nucleon target is concerned, the two-photon exchange
mechanism can be constrained experimentally from parity-conserving
observables both from electron vs.\ positron scattering comparisons and
from dedicated polarization measurements \cite{ABCCV}. 
These data can further be used to provide necessary input for
theoretical models needed to evaluate additional contributions from the
$\gamma Z$-box.

In summary, we have demonstrated that the two-photon exchange
corrections to the parity-violation electron scattering asymmetries can
reach a few per cent, $i.e.$, they are comparable in size to
existing experimental measurements \cite{PVES_exp,Beise03} on the contribution
of strange quarks to these asymmetries.
    
%%%%%%%%%%%%%%%%%%%%%%%%%%%%%%%%%%%%%%%%%%%%%%%%%%%%%%%%%%%%%%%%%%%%%%%%%

%%%%%%%%%%%%%%%%%%%%%%%%%%%%%%%%%%%%%%%%%%%%%%%%%%%%%%%%%%%%%%%%%%%%%%%%%

\begin{acknowledgments}
We ackowledge useful discussions with D. Beck, E.~Beise, T.W. Donnelly, B. Holstein,
K. Kumar, D.~Mack and P. Souder.
This work was supported by the US Department of Energy
under contract DE-AC05-84ER40150 (A.V.A.) and by the National Science Foundation
under grant PHY-0245056 (C.E.C.)
\end{acknowledgments}

%%%%%%%%%%%%%%%%%%%%%%%%%%%%%%%%%%%%%%%%%%%%%%%%%%%%%%%%%%%%%%%%%%%%%%%%%

\end{document}